\documentclass[a4paper]{jpconf}
\usepackage{graphicx}
\usepackage{amsmath}
\let\footnotesize=\scriptsize
\makeatletter
\def\@fnsymbol#1{\ifnum\thefootnote=99\hbox{$^*\m@th$}\else^{\thefootnote}\fi\relax}
\def\blfootnote{\xdef\@thefnmark{}\@footnotetext}

\makeatother

\newif\ifpreprint
\preprinttrue

\def\BlackHat{\textsc{BlackHat}}
\def\Sherpa{{\rm SHERPA}}
\def\Comix{\rm COMIX}
\def\FastJet{\textsc{FastJet}}
\def\Root{\textsc{Root}}
\def\ntuple{$n$-tuple}

\def\NTuple{$N$-Tuple}
\def\ntuples{$n$-tuples}

\def\pT{p_{\rm T}}

\def\Wjn{$W\,\!+\,n$}
\def\Wjnm{$W\!+\,(n\!-\!1)$}

\def\Wj{$W\,\!+\,1$}
\def\Wjj{$W\,\!+\,2$}
\def\Wjjj{$W\,\!+\,3$}
\def\Wjjjj{$W\,\!+\,4$}
\def\Wjjjjj{$W\,\!+\,5$}
\def\Wjjjjjj{$W\,\!+\,6$}
\def\Wmjjjjj{$W^-\,\!+\,5$}

\begin{document}
\ifpreprint
\fi
\title{%
\ifpreprint
{\rm\small
UCLA/13/TEP/109 $\null\hskip 2 cm \null$\hfill
SB/F/423-13$\null\hskip 2 cm \null$\hfill
SLAC-PUB-15760}\break
{\rm\small
IPPP-13-85  $\null\hskip 2 cm \null$ \hfill
Saclay--IPhT--T13/229}\break
\break
\fi
The BlackHat Library for One-Loop Amplitudes}

\author{Z.~Bern$^1$, L.~J.~Dixon$^2$, F.~Febres~Cordero$^3$, 
S.~H{\"{o}}che$^4$, H.~Ita$^5$, D.~A.~Kosower$^{6,}$\footnote[99]{\hskip -1.5mm Presenter\ifpreprint{} at ACAT 2013, Beijing, China, May 16--21, 2013\fi}, 
D.~Ma{\^{i}}tre$^7$, K.~J.~Ozeren$^8$}

\address{
$^{1,8}$Department of Physics and Astronomy, UCLA, Los Angeles, CA 90095-1547, USA}

\address{
$^{2,4}$SLAC National Accelerator Laboratory, Stanford University, Stanford, CA 94309, USA}

\address{
$^3$Departamento de F\'{\i}sica, Universidad Sim\'on Bol\'{\i}var,  Caracas 1080A, Venezuela}

\address{
$^5$Physikalisches Institut, Albert-Ludwigs-Universit\"at Freiburg,
D--79104 Freiburg, Germany}

\address{
$^6$Institut de Physique Th\'eorique, CEA--Saclay,
F--91191 Gif-sur-Yvette cedex, France}

\address{
$^7$Institute for Particle Physics Phenomenology, University of Durham, Durham DH1 3LE, UK
}

\begin{abstract}
We present recent next-to-leading order (NLO) 
results in perturbative QCD obtained
using the \BlackHat{} software library.  We discuss the use of \ntuples{}
to separate the lengthy matrix-element computations from 
the analysis process.
The use of \ntuples{} allows many analyses to be carried out on the
same phase-space samples, and also allows experimenters to conduct their
own analyses using the original NLO computation.
\end{abstract}

\section{Introduction}

Experimental studies of Standard-Model processes at the Large Hadron
Collider (LHC), whether as standard candles, as backgrounds to new
physics, or as part of a program of precision measurements, require
a theoretical counterpart.  The dominant contributions to theoretical
predictions arise from perturbative quantum
chromodynamics (QCD).  In QCD, unlike the case
for other components of the Standard Model,
leading-order (LO) calculations, which rely on tree-level
amplitudes in QCD, do not offer a quantitatively reliable prediction.
This problem is a result of the uncompensated
dependence of the prediction on the
renormalization and factorization scales.  At LO this dependence arises
only via the running coupling and the evolving parton distribution
functions.  Next-to-leading order (NLO) calculations in QCD cure
this problem, and are thus the
the first order in perturbation theory to provide a quantitatively
reliable estimate of backgrounds due to Standard-Model processes~\cite{LesHouchesI,LesHouchesII}.

Recent years have seen remarkable progress in NLO calculations, thanks
to advances in computing the one-loop matrix elements which lie at
their core~\cite{UnitarityMethod, Zqqgg, NewUnitarity,GenHel,DDimUnitarity,
Forde,OPP,Badger,CutTools,BlackHat,Codes,Recent}.  
The \BlackHat{} collaboration has
developed its eponymous software~\cite{BlackHat} to implement 
computations of one-loop amplitudes in a numerical approach.  
For multiplicities beyond \Wjj-jet production, essentially only
tree-level amplitudes and integrals are implemented in computer
code directly from analytic formulas.  The library implements
so-called on-shell methods, which rely only on information present
in amplitudes with on-shell external states in order to arrive at
values for more complex amplitudes.  This keeps the relative
simplicity --- compared to the factorial complexity one might have
feared from a Feynman-diagrammatic expansion --- manifest at every
stage of a calculation.  This simplification is particularly 
evident in high-multiplicity calculations, for which the frontier has
moved from \Wjj-jet production~\cite{Zqqgg,OtherZpppp,MCFM}
 a decade and a half ago to \Wjjjjj-jet 
production earlier this year~\cite{W5j}.

\section{\NTuple{}s}

While numerical one-loop libraries have made previously-unattainable
NLO calculations feasible, they are not the only ingredient needed.
In addition to the tree-level matrix elements for the basic process,
we also need tree-level matrix elements with an additional parton
emission: a gluon emitted, or a gluon splitting to a quark--antiquark
pair.  We use the \Comix~\cite{Comix} library within
\Sherpa~\cite{Sherpa} to compute these matrix elements.  Different
contributions to an NLO calculation have infrared divergences: in the
virtual corrections, they are present explicitly, in the form of poles
in the dimensional regulator $\epsilon$; in the real-emission
corrections, they arise in phase-space integrals over squared matrix
elements.  These divergences cancel in physical quantities such as
cross sections and distributions of infrared-safe observables.  In
order to make this cancellation manifest in an operationally-tractable
manner, one subtracts local approximations to the squared
real-emission matrix elements, and adds back the analytic integrals of
these local approximations.  These integrals have manifest
singularities, again in the form of poles in $\epsilon$, which cancel
those in the virtual corrections.  The local subtraction terms and the
computation of the appropriate subtraction integrals (using analytic
formulas) are also performed by the \Comix{} library.  We thus have
four types of contributions to integrate numerically: Born (B), virtual
corrections (V), integrated subtraction (I), and subtracted real
emission (R).

A typical
calculation at high multiplicity will involve dozens of subprocesses
(two-quark, four-quark, six-quark, etc.{} in all possible crossings).
Each must be integrated over final-state phase space, subject
to cuts simulating the experimental ones.  We use \Sherpa{}
to manage the subprocesses and perform the required 
numerical integrations.
We use the \FastJet{} library~\cite{FastJet}
 to apply a jet algorithm to phase-space
configurations of partons.  The different contributions span a
large numerical range; to focus computer time on the largest
contributions, it is helpful to further split up each of the
four types (B,V,I,R) into parts, corresponding to leading and
subleading color, or to different initial states, etc.

This subdivision does exact a penalty, in the form of the increased
complexity of managing a calculation, tracking all the different 
parts, and ensuring that each part was run correctly.
In a physics study conducted using stand-alone NLO codes such
as {\tt NLOJET++\/}~\cite{NLOJet}, one would write code for a set of
observables, and obtain distributions in those observables by
running the complete code which computes both squared matrix elements
and observables.  Computing new observables as a physics study evolves;
varying renormalization and factorization scales;
computing uncertainties due to our imprecise knowledge of
parton distribution functions (PDFs), would all require rerunning the code
from scratch.  This would force recomputation of the 
short-distance matrix elements, which are much more expensive to
compute than observables.  It would also force us to confront the
subprocesses' and subparts' management complexity anew each time.

It would simplify the calculation to separate the computation of the
squared matrix elements (along with the generation of phase-space
configurations) from the computation of observables.  We do this by
saving generated phase-space configurations, along with their weights,
in a set of \Root{}~\cite{Root} \ntuple{} files.  The weights are
obtained by combining the squared matrix elements with the phase-space
measure, parton-distribution values, and appropriate Jacobian factors.
In addition, we include a set of auxiliary coefficients which allow
the recomputation of the weights using different renormalization and
factorization scales, or different PDF sets.  The files are quite
voluminous, especially for higher multiplicities, so the compression
offered by \Root{} yields a significant reduction compared to a naive
binary format both in disk-space usage and in transmission times.  A
lightweight analysis code scans the resulting files, and computes
cross sections or histograms of observables.  This code can be rerun
as needed for new observables or for estimating scale-sensitivity of
observables, at relatively low cost in computer time.

The use of \ntuple{} files has a bonus: we can hand over sets of \ntuple{}
files to our experimenter colleagues, who can then run their own analyses
on the events they contain.  This avoids the experimenters having to 
learn how to run a somewhat complicated set-up, and avoids them having
to adopt our code or framework for performing analyses.  The analyses
are of parton-level fixed-order events, so that only distributions 
[of infrared-safe observables] can
be compared in a meaningful way to experimental data.  (The \ntuples{}
we have computed to date are not suitable as input to existing
NLO-matched parton-shower codes.)

\section{Vector Boson+Jets}

\begin{figure}[ht]
\includegraphics[clip,scale=0.6]{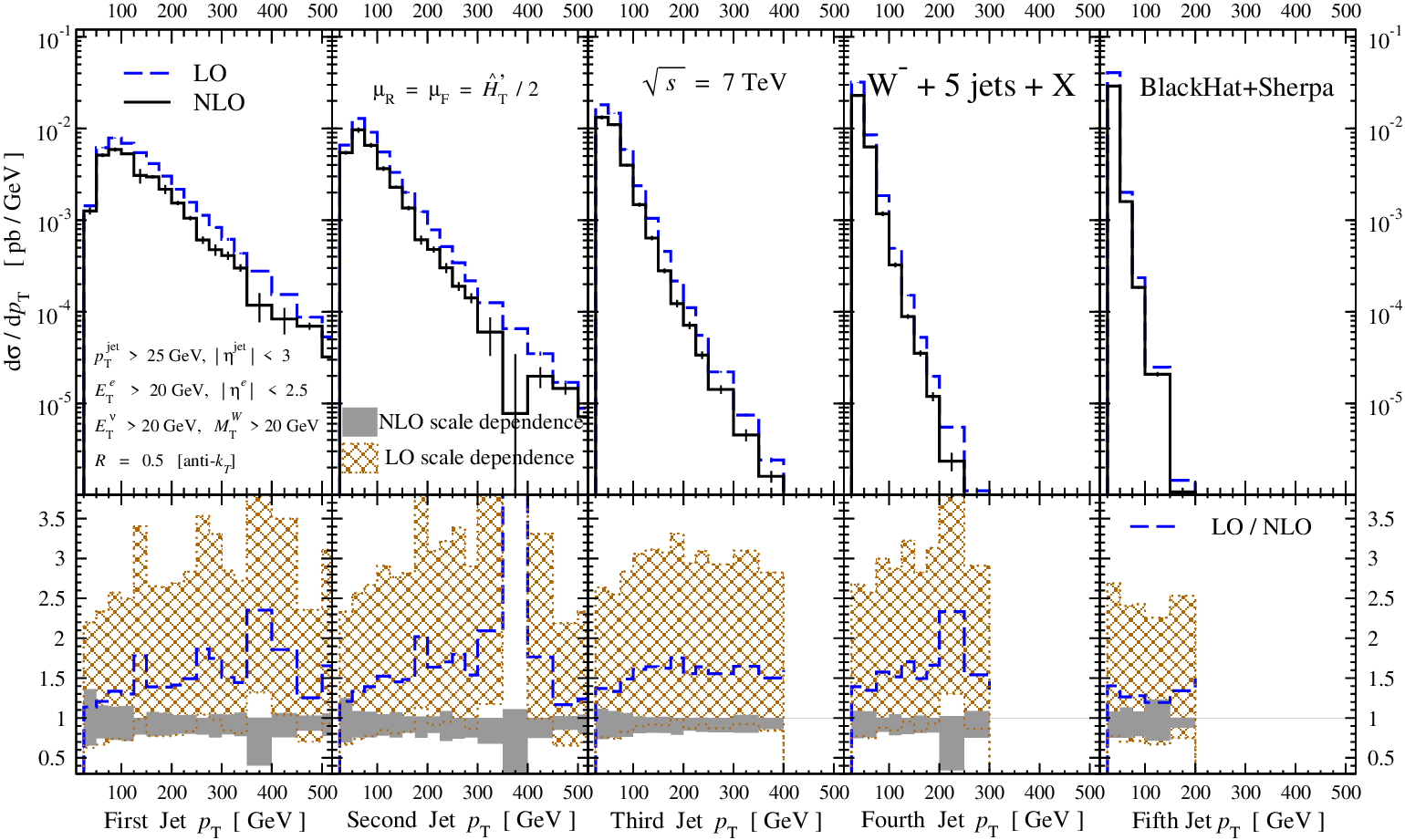}
\caption{\label{W5jFigure} 
The $\pT$ distributions of the leading five jets in \Wmjjjjj-jet production,
ordered in decreasing $\pT$ from left to right.  In the upper panels,
the blue (dashed) line shows the LO prediction, and the black (solid) line,
the NLO one.  The lower panels show the ratio of the central LO prediction
to the NLO one, along with the LO scale variation band in orange-brown 
(hatched) and the NLO scale variation band in gray (solid).
}
\end{figure}

As an example of recent studies we have performed using the \BlackHat{}
library running under \Sherpa{}'s control, and using \ntuple{} files, 
we show the distribution of the leading five jets in 
an NLO computation~\cite{W5j} 
of \Wmjjjjj-jet$\,\!+X$ production in fig.~\ref{W5jFigure}.  
Successive jet
distributions fall more steeply.  This reflects the fact that moving out
on the distribution of, say, the third jet allows the fourth and fifth
jet to remain at low $\pT$, thus forcing up the center-of-mass energy
of the short-distance partonic collision less than moving out on the
distribution of the fifth jet.  As the center-of-mass energy grows, the
short-distance cross section decreases rapidly, both because of direct
dependence in the squared matrix element and flux factors, and because
the parton distributions decrease as well.

For the leading through next-to-softest jet (the fourth jet in this case),
the NLO distributions fall somewhat more steeply, as reflected in the
upward tilt of the LO/NLO ratio in the bottom panels of fig.~\ref{W5jFigure}.
This continues a pattern seen at lower multiplicity,
for example in the \Wjjjj-jet calculation~\cite{W4j}.

The scale-variation bands shrink dramatically in the NLO calculation
as compared to the LO one.  This fulfills one of the primary
motivations for pursuing this calculation: we have achieved a 10--15\%
level of reliability, within expected experimental uncertainties for
this high-multiplicity process.

\begin{figure}[h]
\hskip 15mm 
\includegraphics[clip,scale=0.6]{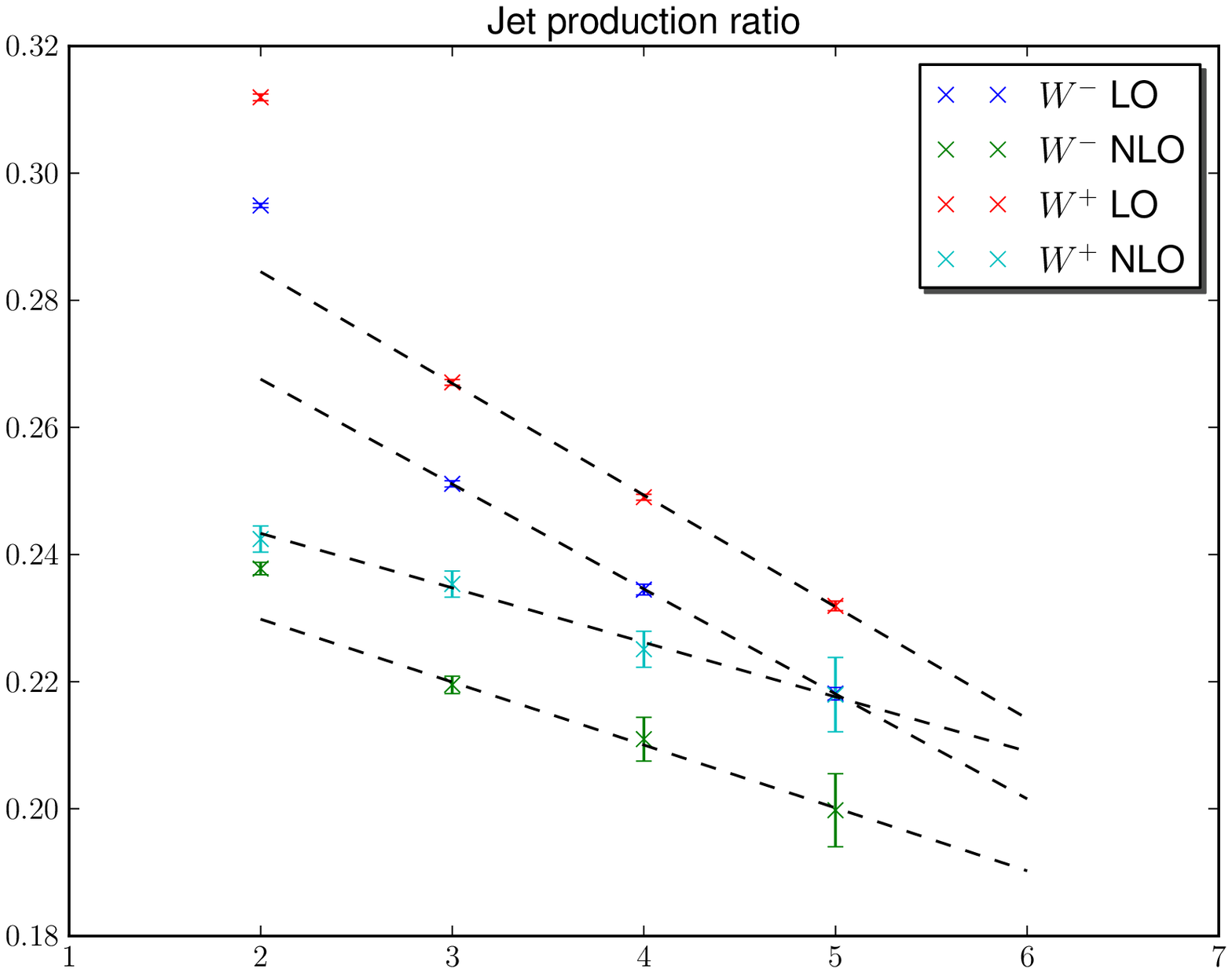}
\caption{\label{RatioFigure} 
Ratios of \Wjn- to \Wjnm-jet production at LO and NLO.  The uppermost
set of points shows the ratios for $W^+$ production at LO; the next
set, for $W^-$ production at LO; the third, for $W^+$ production at NLO;
and the bottommost, $W^-$ production at NLO.
}
\end{figure}
There is an added bonus, if we consider ratios of cross sections
for \Wjn- to \Wjnm-jet production.  This ``jet-production'' 
ratio~\cite{JetProductionRatio} is sometimes called the ``Berends'' or 
``staircase'' ratio.  
We should expect the \Wj-jet
cross section to behave quite differently from cross sections with
higher multiplicity, because some subprocesses (for example, $gg$-initiated
ones) are missing at LO, and also because of kinematic constraints
on phase space both at LO and at NLO.  This in turn means that the
\Wjj- to \Wj-jet ratio should be expected to behave quite differently
than higher-multiplicity ratios.  There are also kinematic constraints evident
in \Wjj-jet production at LO, but these are largely relaxed at NLO;
so we might expect to see signs of regularity or universality appearing
with the \Wjjj- to \Wjj-jet ratio at NLO.  
We consider the ratios of total cross sections with
the cuts as shown in the leftmost panel of fig.~\ref{W5jFigure}.
If we plot these ratios,
as shown in fig.~\ref{RatioFigure},
regularity is exactly what we see: the ratios for \Wjjj- to \Wjj-jet,
\Wjjjj- to \Wjjj-jet, and \Wjjjjj- to \Wjjjj-jet production all lie
on straight lines, both at LO and NLO.  The former is a bit surprising,
but it is the latter that is most significant for applications to
LHC physics.  It allows us to extrapolate linearly, and make a prediction
for \Wjjjjjj-jet production.  In this extrapolation, the \Wjjjjj-jet
calculation plays a crucial role in establishing the reasonableness of
a linear extrapolation.  We find (see ref.~\cite{W5j} for details),
\begin{eqnarray}
&&W^- + 6 \hbox{ jets}:\quad 0.15 \pm 0.01\ \hbox{pb}\,,\nonumber\\
&&W^+ + 6 \hbox{ jets}:\quad 0.30 \pm 0.03\ \hbox{pb}\,,
\end{eqnarray}
using the same cuts as shown in fig.~\ref{W5jFigure}.  The uncertainties
are estimated using a Monte-Carlo procedure with synthetic data sets.

\section*{Acknowledgments}

This research was supported by the US Department of Energy under
contracts DE--AC02--76SF00515 and DE--FG02--13ER42022.  DAK's research
is supported by the European Research Council under Advanced
Investigator Grant ERC--AdG--228301.  
DM's work was supported by the
Research Executive Agency (REA) of the European Union under the Grant
Agreement number PITN--GA--2010--264564 (LHCPhenoNet). SH's
work was partly supported by a grant from the US LHC Theory
Initiative through NSF contract PHY--0705682.  This research used
resources of Academic Technology Services at UCLA, and of the National
Energy Research Scientific Computing Center, which is supported by the
Office of Science of the U.S. Department of Energy under Contract
No.~DE--AC02--05CH11231.

\end{document}